\documentclass[useAMS,usenatbib]{mn2e}

\usepackage{graphicx}
\usepackage{amsmath}
\usepackage{amssymb}
\usepackage{amsbsy}

\newcommand{\MNRAS}{MNRAS}

\newcommand{\rmxaa}{Revista Mexicana de Astronomia y Astrofisica}

\newcommand{\apj}{ApJ}
\newcommand{\apjl}{ApJ}

\newcommand{\aap}{A \& A}

\newcommand{\mnras}{MNRAS}

\newcommand{\nat}{Nature}

\renewcommand{\v}{\ensuremath{\mathbf{v}}}

\newcommand{\B}{\ensuremath{\mathbf{B}}}

\newcommand{\ii}{\ensuremath{\text{i}}}
\renewcommand{\d}{\ensuremath{\partial}}

\newcommand{\ey}{\ensuremath{\mathbf{e}_{y}}}
\newcommand{\ez}{\ensuremath{\mathbf{e}_{z}}}

\makeatletter
\newcommand*{\rom}[1]{\expandafter\@slowromancap\romannumeral #1@}
\makeatother

\title[MRI in debris discs]{The magnetorotational instability in
  debris-disc gas}
\author[Q. Kral, H. Latter]{Quentin Kral$^{1}$\thanks{E-mail: qkral@ast.cam.ac.uk} \& Henrik Latter$^{2}$ \\
$^{1}$Institute of Astronomy, University of Cambridge, Madingley Road, Cambridge CB3 0HA, UK\\
$^{2}$DAMTP, University of Cambridge, CMS, Wilberforce Road, Cambridge CB3 0WA, UK}

\begin{document}

\date{Accepted 1928 December 15. Received 1928 December 14; in original form 1928 October 11}

\pagerange{\pageref{firstpage}--\pageref{lastpage}} \pubyear{2002}

\maketitle

\label{firstpage}

\begin{abstract}
Debris discs are commonly swathed in gas which can
be observed in UV, in fine structure lines in FIR, and in resolved maps of CO
emission. Carbon and oxygen are overabundant in such gas, but it is severely
depleted in hydrogen. As a consequence, its ionisation fraction is
remarkably high, suggesting magnetohydrodynamic (MHD) processes may
be important. In particular, the gas may be subject
to the magnetorotational
instability (MRI), and indeed recent modelling of
 $\beta$ Pictoris requires an
anomalous viscosity to explain the gas's observed radial structure. 
In this paper we explore the possibility that the MRI is active in
debris-disc gas and responsible for the observed mass transport. 
We find that non-ideal MHD and dust-gas interactions play a
subdominant role, 
and that
linear instability is viable at certain radii. However, owing to low gas
densities, the outer parts of the disc could be stabilised by a weak
ambient magnetic field, though it is difficult to constrain such a field.
Even if the MRI is stabilised by too strong a field, 
a magnetocentrifugal wind may be launched in its place and this could
lead to equivalent (non-turbulent) transport. Numerical simulations of the vertically
stratified MRI in conditions appropriate to the debris disc gas should
be able to determine the nature of the characteristic behaviour at
different radii, and decide on the
importance of the MRI (and MHD more generally) on the evolution of 
these discs.

\end{abstract}


\section{Introduction}

Observations of gas in debris discs are becoming increasingly common
thanks to the analysis of UV absorption lines (for edge-on discs),
detections of C II and O I by Herschel,
and highly resolved sub-mm images of CO and the continuum provided by ALMA
 \citep[e.g.][]{2014Sci...343.1490D}.
One of the more important discoveries is that gas in these old systems
is not a remnant of their preceding protoplanetary discs, but is
continually replenished by the disc's solid bodies,
either by photodesorption \citep{2007A&A...475..755G}
or grain-grain collisions \citep{2007ApJ...660.1541C}.

Amongst debris discs with gas, $\beta$ Pic is the best constrained. 
Molecules such as CO \citep{2014Sci...343.1490D}, atoms such as C,
O, S, and a variety of metals have been detected in 
$\beta$ Pictoris' gas disc \citep{2006Natur.441..724R}. 
Hydrogen, on the other hand, is thought to be highly depleted 
\citep{1995A&A...301..231F,2001Natur.412..706L}, and in fact 
the gas mostly composed of carbon
\citep[e.g.][]{2014ApJ...796L..11R,2014A&A...563A..66C}. 
Partly as a result, the ionisation fraction is exceptionally high: 
in $\beta$ Pictoris, it may be close to 0.5 \citep{2014A&A...563A..66C}, significantly
greater than a protoplanetary
disc, though gas densities in debris discs are much less.

Recently, \citet{kral16} developed a new model to study the evolution
of gas in debris discs.
They propose that (a) CO is created within the debris belt (by grain
collisions or photodesorption), (b) quickly 
photodissociated into carbon and oxygen, and then (c) that the two atomic elements spread
radially, due to an anomalous viscosity that may be parameterised
with the $\alpha$ prescription \citep{1973A&A....24..337S}. 
The resulting time and radius dependent ionisation fractions,
temperature, and number densities of carbon and oxygen 
are computed using the PDR-like algorithm, Cloudy
\citep{2013RMxAA..49..137F}, 
which is coupled to a 1D dynamical $\alpha$ model for the disc structure.
In order to reproduce the gas observations in $\beta$ Pic,
\citet{kral16} require the diffusion of gas to be very effective. 
In other words, there must be significant turbulent activity in
order to transfer enough angular momentum. In fact, they predict that
$\alpha > 0.1$ in the gas disc. 
What could be the cause of such a high level of turbulence? 
Given the typical ionisation fractions, could this transport be the result
of the magnetorotational instability
\citep[MRI,][]{1998RvMP...70....1B}?
It is this question that our paper is concerned with.

Certainly, the highly ionised discs associated with close binaries yield
$\alpha$ values ranging from 0.1 to 0.4 in their high states
\citep{2007MNRAS.376.1740K}, while the decretion discs orbiting Be
stars sometimes exhibit even higher values
\citep{2012ApJ...744L..15C}.
As the MRI is thought to be responsible for the transport in these
systems, so too could it drive gas diffusion in debris discs. 
Numerical simulations of the MRI also report $\alpha$ values that can
reach up to 0.1
\citep[e.g.][]{2004ApJ...605..321S,2009ApJ...690..974S,2016MNRAS.457..857S}. 
However, the physical conditions
in debris-disc gas is potentially different to these classical
applications. Of greatest concern is the gas's low density, which make
it vulnerable to ambipolar diffusion and the stabilising role of
magnetic tension. 

In this paper, we check the linear stability of the gaseous envelope
of a debris disc to the MRI, focussing on the system 
that is best characterised: $\beta$ Pictoris. We use the model
presented in \citet{kral16} to determine the parameters of its gas disc
and then determine its linear stability, following the formalism of
\citet[][hereafter KB04]{2004MNRAS.348..355K}. We find that, on the whole,
non-ideal MHD effects are not crucial, nor is the dust-gas coupling.
However, relatively weak fields
can stabilise the MRI; at 100 AU, fields greater than a $\mu$G are
sufficient to deactivate the fastest growing modes (channel flows), though much slower
radial-modes can still function, presumably leading to only weak
transport. It is possible that
at these larger radii a magnetocentrifugal wind supplants the MRI and
provide the effective $\alpha$ that observations and modelling
demand \citep{2012MNRAS.423.1318O,2013ApJ...769...76B,2013A&A...550A..61L}. We do not examine in detail this
 and alternative nonlinear outcomes, but future numerical work in
 vertically stratified boxes tailored to debris disc conditions
 will be able to determine how the gas radially spreads.

In Section 2, we present the state-of-the-art of gas modelling in debris discs
and give some details of the physical regime in which $\beta$ Pictoris
lies. 
In Section \ref{maths} we present the linear theory of the ambipolar
MRI and test it on $\beta$ Pic. Finally,
we discuss these results, previous simulation results that are
relevant, and point to future work.

\section{Debris disc properties}\label{debris}

\subsection{Debris disc gas}

The gas that cocoons debris discs appears
depleted in hydrogen but abundant in carbon and oxygen
\citep{2010ApJ...720..923Z,2014A&A...563A..66C}. 
These discs are observed around young main sequence stars where the
main ionising fluxes 
come from the interstellar radiation field (IRF) and the star itself. It is worth
pointing out that the main source of ionisation in protoplanetary
discs (X-rays produced by stellar coronal activity) is
absent at this later evolutionary stage. C I is ionised easily by the
IRF (and the stellar radiation field) 
because its ionisation potential (IP) is 11.26eV, whilst hydrogen and oxygen IPs 
are 13.6eV, just above the limit set by the Lyman break. 
One can then regard these discs as mainly composed of C II
(ionisation fraction $\sim$ 0.5), 
C I and O I. Some other metals are observed but in smaller
quantities. If it were not for Coulomb collisions with C II, these
metals would leave the system on a few dynamical timescale due to
radiation pressure \citep{2006ApJ...643..509F,2001ApJ...563L..77O}. We
neglect, however,
these relatively scarce metals.

\citet{kral16} propose that the gas in debris discs is 
created within the main belt by either photodesorption or grain-grain collisions. CO is released from the grains and photodissociates
quickly into carbon and oxygen. Oxygen remains neutral but carbon
becomes ionised. 
The atomic gas then evolves viscously, due to turbulence parametrised by an
$\alpha$ viscosity.
 As the viscosity depends on temperature, \citet{kral16}
 use a PDR-like code (called `Cloudy') 
to follow the thermal evolution of the gas
at the same time as its dynamics. 
The temperature is fixed by a competition between the carbon
photoionisation heating 
and the cooling by the C II fine structure line. The free parameters
in the model are $\alpha$, 
the mass input rate of CO, the location of the debris belt and the amount
of UV flux impinging on the gas disc. The CO input rate
and location of the debris belt are determined by 
 ALMA observations of $\beta$ Pic. Once these are fixed, \citet{kral16} 
find that all other observations are reconciled if $\alpha \gtrsim
0.1$. Thus the anomolous transport required is very efficient.

\begin{figure*}
   \centering
   \includegraphics[width=18cm]{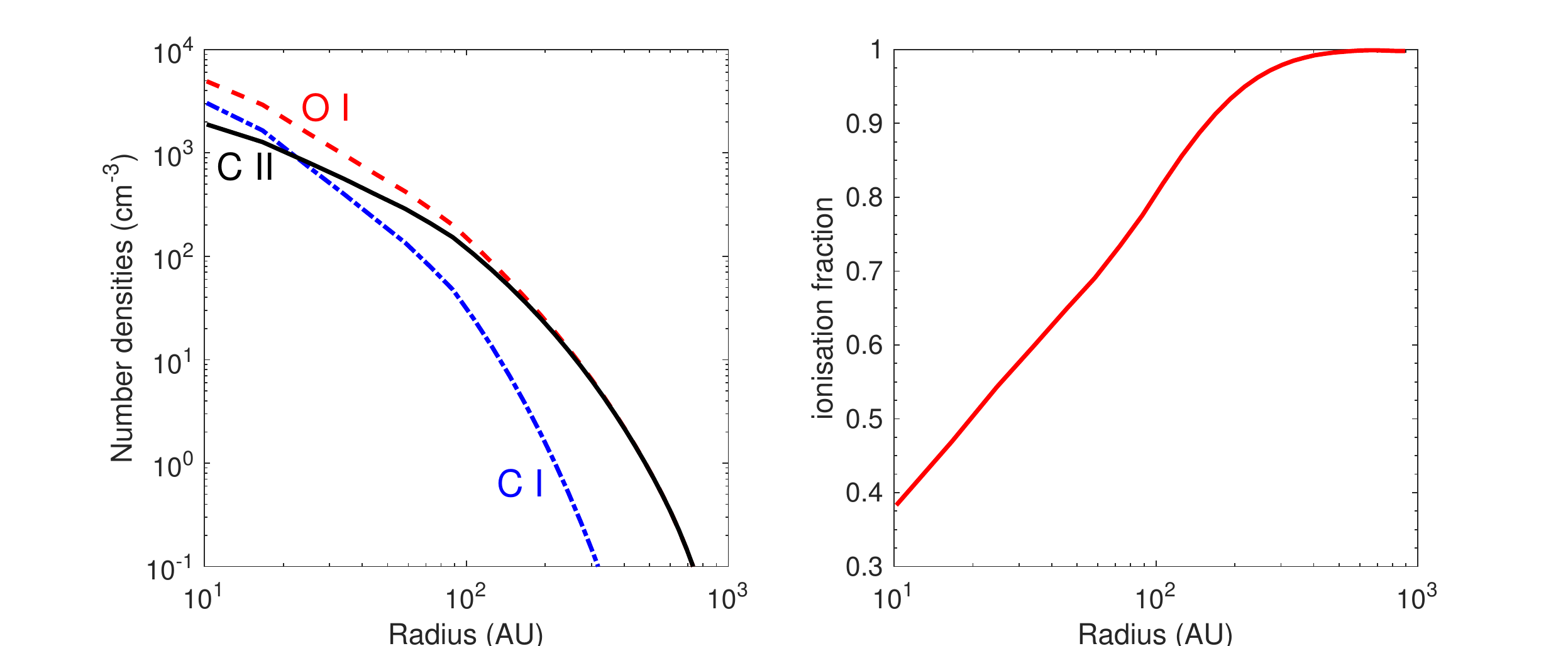}
   \caption{\label{figdens} $\beta$ Pic best-fit model from
     \citet{kral16}. {\it Left:} Densities of C I (dashed-dotted
     blue), C II (solid black) and O I (dashed red). The electron density is the same as the C II density. {\it Right:} Carbon ionisation fraction as a function of $R$.}
\end{figure*}

\begin{figure*}
   \centering
   \includegraphics[width=17cm]{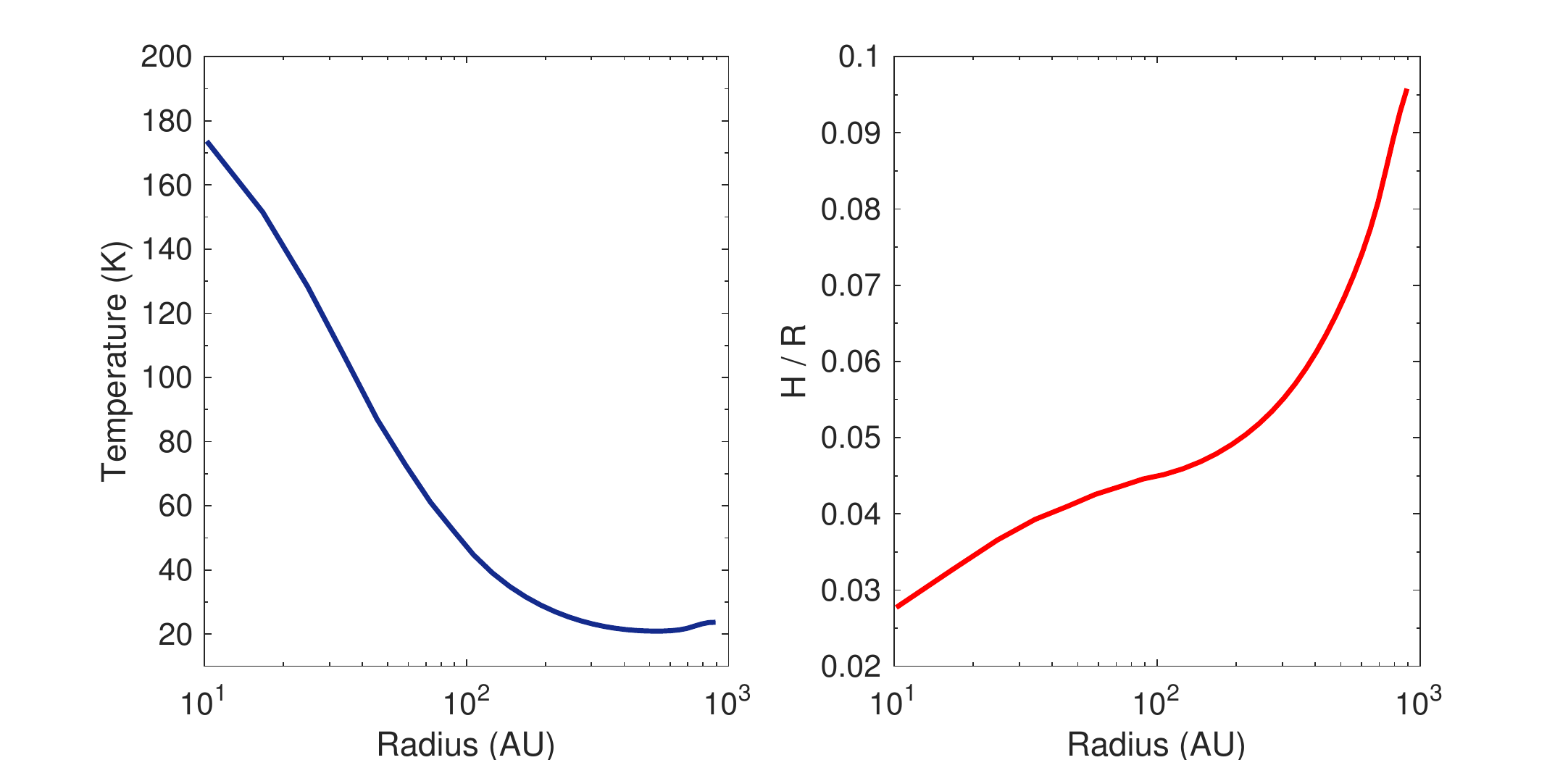}
   \caption{\label{fighsurr} $\beta$ Pic best-fit model from \citet{kral16}. {\it Left:} Temperature as a function of $R$. {\it Right:} $H$ over $R$ as a function of $R$.}
\end{figure*}

In Fig.~\ref{figdens} we show the best-fit model of the $\beta$ Pic
gas disc in \citet{kral16}
to give a sense of the physical regime it lies in and which we explore
in this paper. 
The number densities for carbon and oxygen are lower than $10^4$ cm$^{-3}$. The oxygen density equals the sum of the C I and C II densities
because these atomic elements arise from the breakup of CO. 
Because of the depletion of hydrogen,
 the dynamics of the system and its thermal evolution is controlled by
 the carbon content. One can see that the C II density profile
 flattens in the inner region. 
This is a consequence of the accretion disc structure.
The carbon density reaches higher values in the inner region and so
 the medium becomes optically thick to FUV; as a result, 
carbon ionisation is impeded (see Fig.~\ref{figdens} right). 
The electron density is superimposed on the C II line in Fig.~\ref{figdens} (left) as electrons are produced 
through C I photoionisation. Finally, \citet{kral16} showed that the dust does
 not have any thermal effect on the gas state because the photoelectric
 effect is not sufficiently strong. 

For comparison with a typical protoplanetary disc we employ 
the Minimum Mass Solar Nebula (MMSN) \citep{1981PThPS..70...35H}. 
At 10AU and 100AU, $\beta$ Pic surface densities are $\sim$ 8 and 4$
\times 10^{-7}$ g cm$^{-2}$
 whilst
the MMSN gives $\sim 50$ and 1.7 g cm$^{-2}$. Thus, the debris disc
gas possesses a surface density $10^7$ times lower than typical
protoplanetary discs. 
However, the ionisation fraction is significantly higher, $\gtrsim
0.1$. In contrast, a protoplanetary disc at 10 AU yields 
fractions between $10^{-10}$ (at the
midplane) to $10^{-5}$ in the disc corona \citep{2014A&A...566A..56L}.  
Finally, the ionisation timescale in $\beta$ Pic is found to be on the order of a few years, which
is markedly shorter than the local orbital period everywhere for the
radial range considered in Fig.~\ref{fighsurr}.

The temperature varies between $\sim$ 110 and 20K for $\beta$ Pic, 
but is not a simple power law as shown by Fig.~\ref{fighsurr}
(left). 
Three regimes can be observed as explained in \citet{kral16}. In most of the disc,
the temperature drops as $R^{-0.8}$ (before flattening out) due to the
increasing ionisation fraction of carbon, 
which provides more C II to cool the system. The absolute value of the
temperature and the profile is 
similar to what is expected in protoplanetary discs \citep[e.g.][]{1997ApJ...486..372B}. The derived
$H/R$ is shown on Fig.~\ref{fighsurr} (right) and varies between
$\sim$ 0.03 at 10 AU and 0.1 at 1000 AU. 

\subsection{Non-ideal MHD}

Owing to the low number densities, and the consequently weaker collisional
coupling, one may expect non-ideal MHD effects to raise their
heads. The importance of these effects may be quantified by the
Elsasser numbers for Ohmic diffusion, the Hall effect, and ambipolar
diffusion. The first two are given by
\begin{align}
E_\text{O} = \left(\frac{4\pi\, e^2}{c^2\,
    m_e\,\Omega}\right)\frac{x_e}{\langle \sigma v
  \rangle_{ne} + \langle \sigma v
  \rangle_{ie}}\,v_A^2, \qquad
E_\text{H} = \frac{B\, e\, x_e}{m_n\,\Omega c}
\end{align}
respectively. Here $\Omega$ is the orbital period, $c$ is the speed of
light, $x_e$ is the
electron fraction, $m_e$ and $m_n$ are the masses of electrons and
neutrals, $e$ is charge, $B$ is the imposed magnetic field, and $v_A=
B/\sqrt{4\pi\rho}$ is the Alfv\'en speed (with $\rho$ mass density of the
fluid). Finally, 
 $\langle \sigma v \rangle_{ne} \sim 8.28 \times 10^{-9} \left( T/100 K
 \right)^{0.5}$ cm$^3$ s$^{-1}$ and $\langle \sigma v \rangle_{ie} \sim 0.37 \, \ln{\lambda}/17 \, \left( T/100 K
 \right)^{-1.5}$ cm$^3$ s$^{-1}$, where $\ln{\lambda}$ is the Coulomb logarithm.
Estimates for the two Elsasser numbers are captured by the following
scalings
\begin{align}
E_\text{O} &\sim 1.3 \times 10^{6}\left(\frac{T}{100\, \text{K}}\right)^{5/2}
\left(\frac{R}{10\, \text{AU}}\right)^{3/2} \notag\\
& \hskip4cm \beta^{-1} \,x_e \left(\frac{\ln{\lambda}}{17} \right)^{-1}, \\
E_\text{H} &\sim 4.8\times 10^5\left(\frac{T}{100\, \text{K}}\right)^{1/2}
\left(\frac{R}{10\, \text{AU}}\right)^{3/2} \notag\\
& \hskip3cm \left(\frac{n}{100\, \text{cm}^{-3}}\right)^{1/2}
\beta^{-1/2}
x_e,
\end{align}
where $\beta= 2c_s^2/v_A^2$ is the plasma beta. Clearly in the case of
$\beta$ Pic, these quantities are extremely large, and Ohmic and Hall
effects can safely be ignored. 

\begin{figure}
   \centering
   \includegraphics[width=9cm]{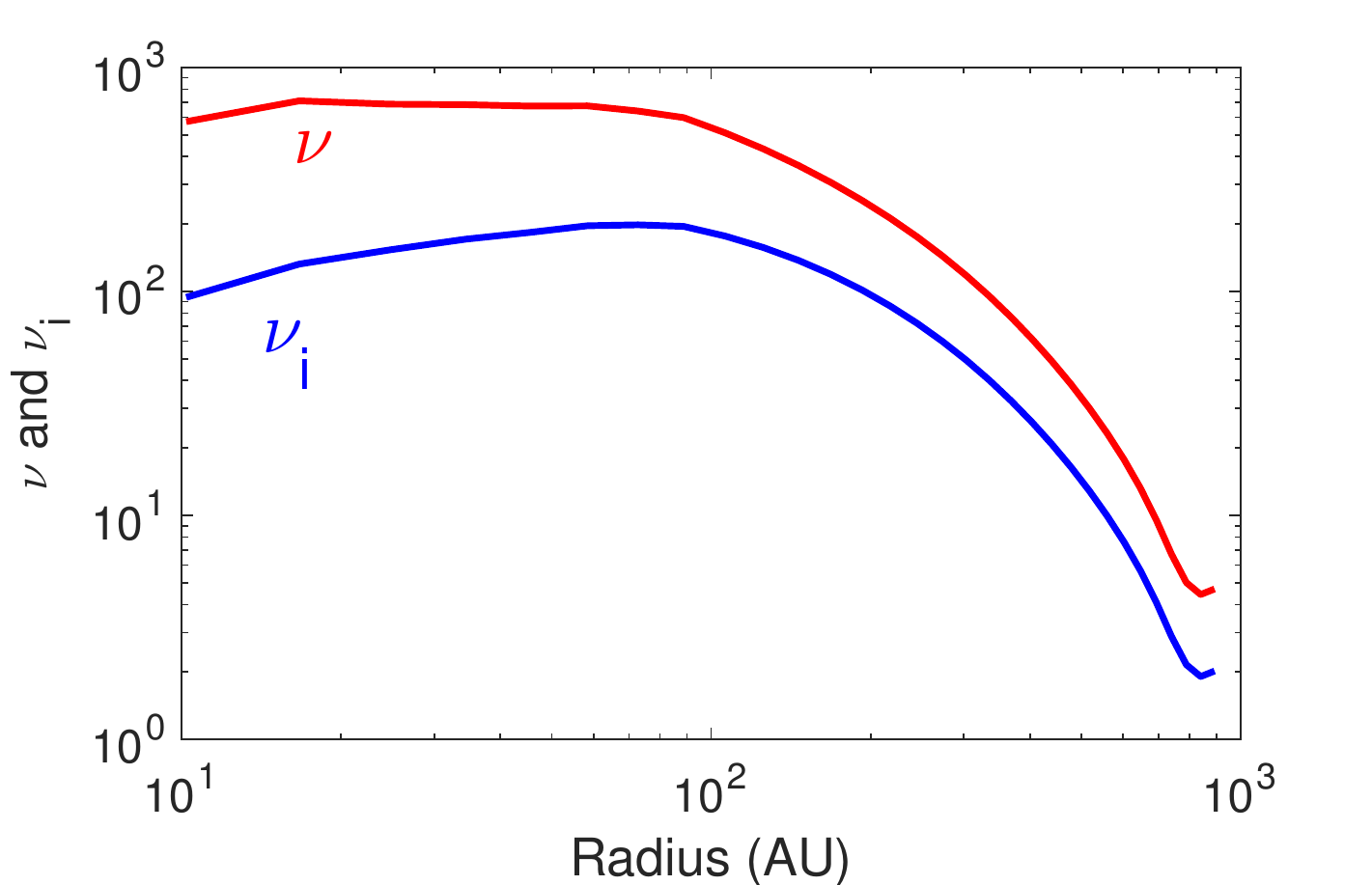}
   \caption{\label{fignu} $\beta$ Pic best-fit model from \citet{kral16}. $\nu$ (red) and $\nu_i$ (blue) as a function of $R$ (see Eq.~\ref{nui}).}
\end{figure}

The Elsasser number for
Ambipolar diffusion, sometimes just called the ambipolar parameter,
is the normalised collision rate between ions and neutral. We
define it via
\begin{equation}
\label{nui}
\nu_i = \frac{\gamma\,\rho_i}{\Omega},
\end{equation}
where $\rho_i$ is the mass density of the ions, and $\gamma$ is the
drag coefficient. It may be calculated via
$\gamma=\langle \sigma v \rangle_i/(m_n+m_i)$,
with $\langle \sigma v \rangle_i = 9.57 \times 10^{-10}$ cm$^3$
s$^{-1}$
 the ion-neutral (C$^+$/O) collision rate
 \citep{2011piim.book.....D}. An equivalent parameter can be computed
 for the total fluid which we denote by $\nu$. Both $\nu_i$ and $\nu$
 are plotted in Figure \ref{fignu}. At most radii both are $\sim 100$,
 and thus neutrals and ions are relatively well coupled. Consequently,
  ambipolar diffusion is not dominant though, as we shall see
 in the next section, its effects are not always negligible. 

\subsection{Dust-gas interactions}

Of course, debris discs are composed of a dusty component that could interfere
with the gas dynamics, and vice versa. In this subsection we explore the coupling
between gas and dust to see whether it could affect the onset of MRI.

The dust number density $n_d$ in $\beta$ Pic varies between
$10^{-11}$ to $10^{-10}$ cm$^{-3}$, from
10 to 200AU \citep{2000ApJ...539..435H,2010ApJ...720..923Z}. The mass
density is less or comparable to that of the gas in the system. As a comparison,
the mass density of the dust in protoplanetary discs
is usually taken to be only 1\% of the gas. Running the model presented in
\citet{kral16}, we find that in general dust grains in debris discs are charged positively with a mean
charge of $\sim 10q$. We also find that the vast majority of electrons
in the gas come from photoionisation of carbon rather than the dust's
photoelectric effect.

As regards the gas's effect on the dust dynamics, 
\citet{kral16} show that gas drag is only important for the smallest
grains, which are scarce in any case because they fall below the
blow-out-size \citep[e.g.][]{2013A&A...558A.121K}. For the most
numerous grain sizes the Stokes number of the dust fluid is $>10$. 

Next we assess the dust's effect on gas dynamics, focusing on $\beta$ Pic.
The neutral-atom/dust collision frequency $p$ can be estimated from
\begin{equation}
\label{eqdust}
\frac{p}{\Omega}=2 \times 10^{-4} \left( \frac{d}{5 \mu \mathrm{m}} 
\right)^2 \left( \frac{n_d}{10^{-10}\mathrm{cm}^{-3}} \right) \left( \frac{T}{50 \mathrm{K}} \right)^{0.5} \left( \frac{R}{10\,\mathrm{AU}} \right)^{1.5},
\end{equation}
\noindent where $d$ is the grain size. In debris discs, the collisional cross-section
is dominated by the smallest
 dust grains close to the blow-out size; the parameters in Eq.~\ref{eqdust} account for that. Even in the worst case scenario 
for grains that are at hundreds of AU, $p/\Omega$ is always smaller
than $10^{-2}$. This estimate indicates there is only weak coupling
between the neutral atoms and dust grains. The two fluids are
mostly free to pursue their own dynamics separately.

However, grains are strongly charged which can increase the
collisional cross-section with ions and electrons.
We used \citet{1987ApJ...320..803D} to quantify the increase in
collisional 
cross section for charged grains. 
The $\epsilon$ variable in \citet{1987ApJ...320..803D} is defined by
 $\epsilon=E\,d/e^2$, where $E$ is the kinetic energy of an ion of
 charge $-e$.
 For charged grains in $\beta$ Pic
it may be estimated by
\begin{equation}
\label{eqeps}
\epsilon=4 \times 10^{4} \left( \frac{d}{5 \mu \mathrm{m}} \right) \left( \frac{R}{10\,\mathrm{AU}} \right)^{-1},
\end{equation}
which is relatively high because of the size of the grains and their
significant charge.
Using this value for $\epsilon$ and the average grain charge, Fig.~1 in
\citet{1987ApJ...320..803D} indicates that the collisional
cross-section is not greatly enhanced. 
Thus the Eq.~\ref{eqdust} is a good estimate for all
constituents of the gas fluid. 

As a consequence, dust drag is neglected in the following section,
though is easy to include and may be relevant for debris discs in
which the dust density is higher. The drag should only influence the
MRI when the collision frequency is $\sim \Omega$. In this case, the
MRI growth rate is decreased, though the instability criterion is not altered
directly, as shown by \citet{2015MNRAS.447.3992G} in the context of protoneutron stars.  

\section{The MRI in debris discs}\label{maths}

In this section we briefly present a MRI stability analysis tailored
to the partially ionised plasma swathing debris discs. The formalism
is lifted from KB04 who include the effects of
ion-neutral drift (ambipolar diffusion) in a single fluid treatment.

As is clear from Figure \ref{fignu}, the coupling timescales of both ions and
neutrals is much less than the orbital timescale throughout most of
the disc, both $\nu$ and $\nu_i$ are $\sim 100$ on radii less than
$\sim 100$ AU.
Thus,
the collisional coupling between the ions and neutrals remains 
sufficiently high for a single-fluid model to be applicable.
For reference, the
outer regions of a protoplanetary disc possess 
$\nu\sim 10^9$ and $\nu_i\sim 10$ \citep[e.g.][]{2011ApJ...739...50B,2015MNRAS.454.1117S}.
Moreover, the electron recombination time is short compared to the
local orbital period.
Ambipolar diffusion plays a relatively minor role, but we retain its
effects in the model. Both the Hall effect and Ohmic
diffusion are omitted. Their respective Elsasser numbers exceed
$10^5$ and thus we deem them to be very much subdominant. 

It should be stressed that gas in other debris discs may be more
dilute and less well ionised, in which case the two-fluid model of
\citet{1994ApJ...421..163B} may be more appropriate. This will certainly be
the case if equilibrium ionisation chemistry is
included: as those authors show, an increasing azimuthal field
markedly reduces
the MRI growth rates, and indeed complete
stabilisation is possible. Note that $\beta$ Pic is too dense for
this effect to be important.

\subsection{Governing equations}

We assume the MRI instigates relatively slow motions 
and excites relatively short scales. We thus adopt the equations of
incompressible non-ideal MHD and a local model of the disc \citep[the
shearing box;][]{1965MNRAS.130...97G}. The latter is a
corotating Cartesian representation of a small `block' of disc fluid
at a given radius $R_0$ orbiting at a rate $\Omega$. Its radial and azimuthal
coordinates are denoted by $x$ and $y$, respectively.

The equations controlling the evolution of the neutral fluid are
\begin{align}
\d_t \v + \v\cdot\nabla \v &= -\frac{1}{\rho}\nabla p -\nabla\Phi_t -2\Omega \ez\times\v 
 +\frac{(\nabla\times\B)\times\B}{4\pi\rho}, \\
\d_t \B + \v\cdot\nabla\B &= \B\cdot\nabla\v + 
\nabla\times\left[\frac{(\mathbf{J}\times\B)\times\B}{c\gamma\rho_i\rho}\right],
\end{align}
where $\v$, $\rho$, and $p$ are the neutral fluid's velocity, density,
pressure, while $\B$ is the
magnetic field. The tidal potential is given by $\Phi_t=
-(3/2)\Omega^2 x^2$, and the current density by
$\mathbf{J}=(c/4\pi)\nabla\times\B$. The ion density is denoted by
$\rho_i$ and $c$ is the speed of light. These equations must
be solved alongside the solenoidal restrictions,
$\nabla\cdot\v=\nabla\cdot \B=0$.

\subsection{Dispersion relation}

We assume for simplicity a homogeneous equilibrium of simple Keplerian
rotation so that $\rho$ is a constant, and
$\v=-(3/2)\Omega x \ey$. In addition, 
a weak vertical and azimuthal field permeates the plasma, so that
$\B=B_y\ey + B_z \ez$. 

We perturb this equilibrium with disturbances of the form 
$\propto \text{e}^{\ii k_x x + \ii k_z z+s t}$, where $k_x$
and $k_z$ are real vertical wavenumber
and $s$ is a (complex) frequency. We omit the details of the
resulting linearised equations, but present the fourth-order dispersion
relation that ensues. For further details the reader may consult \citet{2004MNRAS.348..355K}. The relation, though imposing, can be made more
workable by adopting units so that $\Omega=1$ and $v_{Az}^2 = B^2_z/(4\pi\rho)=1$.
We then obtain
\begin{equation}
s^4 + c_3 s^3 + c_2 s^2 + c_1 s + c_0 =0.
\end{equation}
The four coefficients are
\begin{align*}
& c_3 =k_z^2(\delta^2\varepsilon^{-2}+1)\nu_i^{-1}, \\
& c_2 = \varepsilon^{2} + 2 k_z^2 + \delta^2  k_z^4\varepsilon^{-2}\nu_i^{-2} +
\frac{3}{2} \left(\frac{B_x}{B_z}\right)\left(\frac{k_x}{k_z}\right)k_z^2\nu_i^{-1}, \\ 
& c_1 = c_3(k_z^2 + \varepsilon^{2}), \\
& c_0 = k_z^2(k_z^2 -3\varepsilon^{2}) +
\delta^{2}k_z^4\nu_i^{-2} \notag \\
& \hskip3cm +\frac{3}{2}(k_z^2 + \varepsilon^{2})\left(\frac{k_x}{k_z}\right)
\left(\frac{B_y}{B_z}\right)k_z^2\nu_i^{-1},
\end{align*}
where 
$$ \varepsilon^2= \left(1+\frac{k_x^2}{k_z^2}\right)^{-1}, \qquad \delta^2= 1 +
\frac{B_y^2}{B_z^2}.$$
We may consider $s$ as a function of the vertical wavenumber $k_z$
once the following parameters have been set: $k_x/k_z$, $B_y/B_z$, and
$\nu_i$. 

\subsection{Stability criterion}

Marginal stability corresponds to $s=0$, which permits us
to obtain a stability criterion. In the simple case of channel modes,
 $k_x/k_z=0$,
instability occurs on all vertical wavenumbers satisfying
\begin{equation}
k_z^2 < \frac{3\nu_i^2}{\nu_i^2 + \delta^2}.
\end{equation}
In the strongly coupled limit $\nu_i\gg 1$ we recover the classical
ideal stability criterion for a Keplerian disc. But for smaller values of
$\nu_i$, instability only occurs on longer scales. Moreover, the presence of
an azimuthal field ($\delta>1$) further restricts the allowed range of
wavenumbers and, as we shall see in the next section, 
the growth rates also suffer.

For non-zero $k_x$ the stability criterion is a great deal more
complicated (KB04). Instability
occurs when
\begin{equation} \label{criterion}
k_z^2 <
3\varepsilon^2\frac{\nu_i^2-\frac{1}{2}(k_x/k_z)(B_y/B_z)\nu_i}{\nu_i^2
+ \delta^2 + \frac{3}{2}(k_x/k_z)(B_y/B_z)\nu_i}.
\end{equation}
In the limit of $\nu_i\gg 1$, we recover the expected axisymmetric
criterion in ideal MHD, $k_z^2 < 3\varepsilon^2$ \citep[see for e.g.][]{2015MNRAS.453.3257L}. 
For moderate values of $\nu_i$ the range of permitted $k_z$ is
complicated and in fact, as pointed out by \citet{2004MNRAS.348..355K}, the MRI
can grow on a vastly extended range of $k_z$ in a band of negative
$k_x/k_z$ (albeit with a much reduced growth rate). This phenomena
highlights the role of ambipolar diffusion in mitigating the
stabilising influence of magnetic tension on shorter scales. Field
lines
which would ordinarily prevent tethered fluid blobs from drifting
apart via MRI can slip
through them --- and yet concurrently transport sufficient angular
momentum for instability to proceed. Meanwhile, if the magnetic perturbations are
appropriately oriented, then damping is minimised (see discussion
in Desch 2004).  

Finally, unstable modes must be able to fit into the disc. And thus
$k_z > v_{Az}/c_s$, where $c_s$ is the gas sound speed. Combined with
criterion \eqref{criterion}, this yields a maximum vertical 
field strength that instigates growth for a given mode.

\subsection{Application to $\beta$ Pic}

In this section we adopt $\nu_i=100$, and compute MRI 
growth rates versus wavenumber for various values of $k_x/k_z$ and
$B_y/B_z$.
 We also discuss the stability
criterion and give upper limits on the critical ambient magnetic field
above which the MRI switches off.

\begin{figure}
   \centering
   \includegraphics[width=8.5cm]{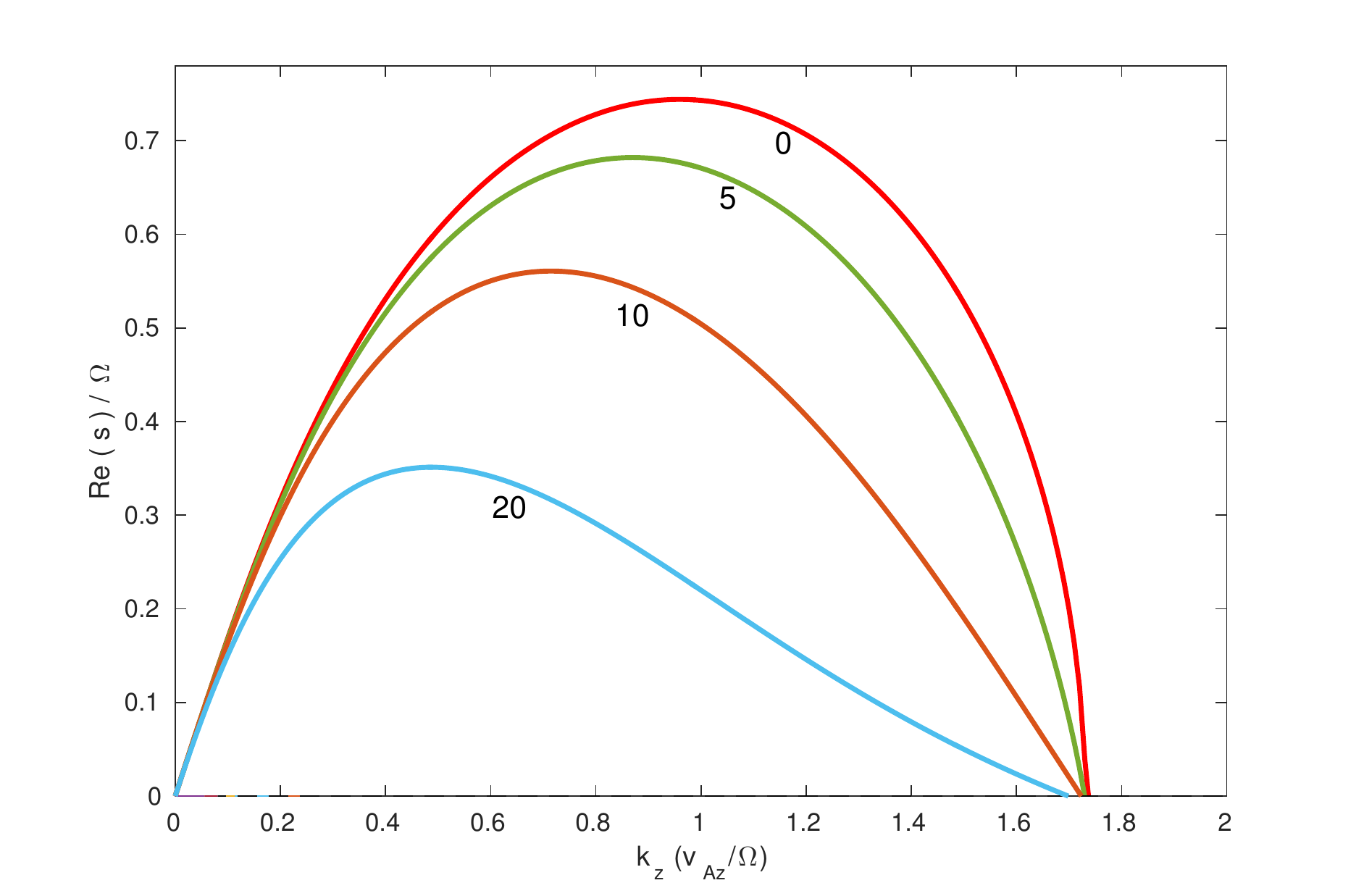}
   \caption{\label{GR1} The real part of the MRI growth rate $s$ as a
     function of vertical wavenumber $k_z$. Modes plotted are channel
     flows, with $k_x=0$, and $\nu_i=100$. 
     The different curves correspond to
     $B_y/B_z=0,\,5,\,10,$ and 20.  }
\end{figure}

In Figure \ref{GR1} we plot the real part of the growth rate for
channel modes, i.e. when $k_x=0$. Four different background
magnetic configurations are presented: $B_y/B_z = 0,\,5,\,10,$ and
$20$. Though the ions and neutrals are relatively well-coupled
($\nu_i=100$), the growth rates deviate from the predictions of ideal
MHD in that the stronger the azimuthal field, the weaker the growth:
an effect clearly associated with ambipolar diffusion. The critical 
wavenumber of marginal stability, however, remains close to the
classical value.

\begin{figure}
   \centering
   \includegraphics[width=8.5cm]{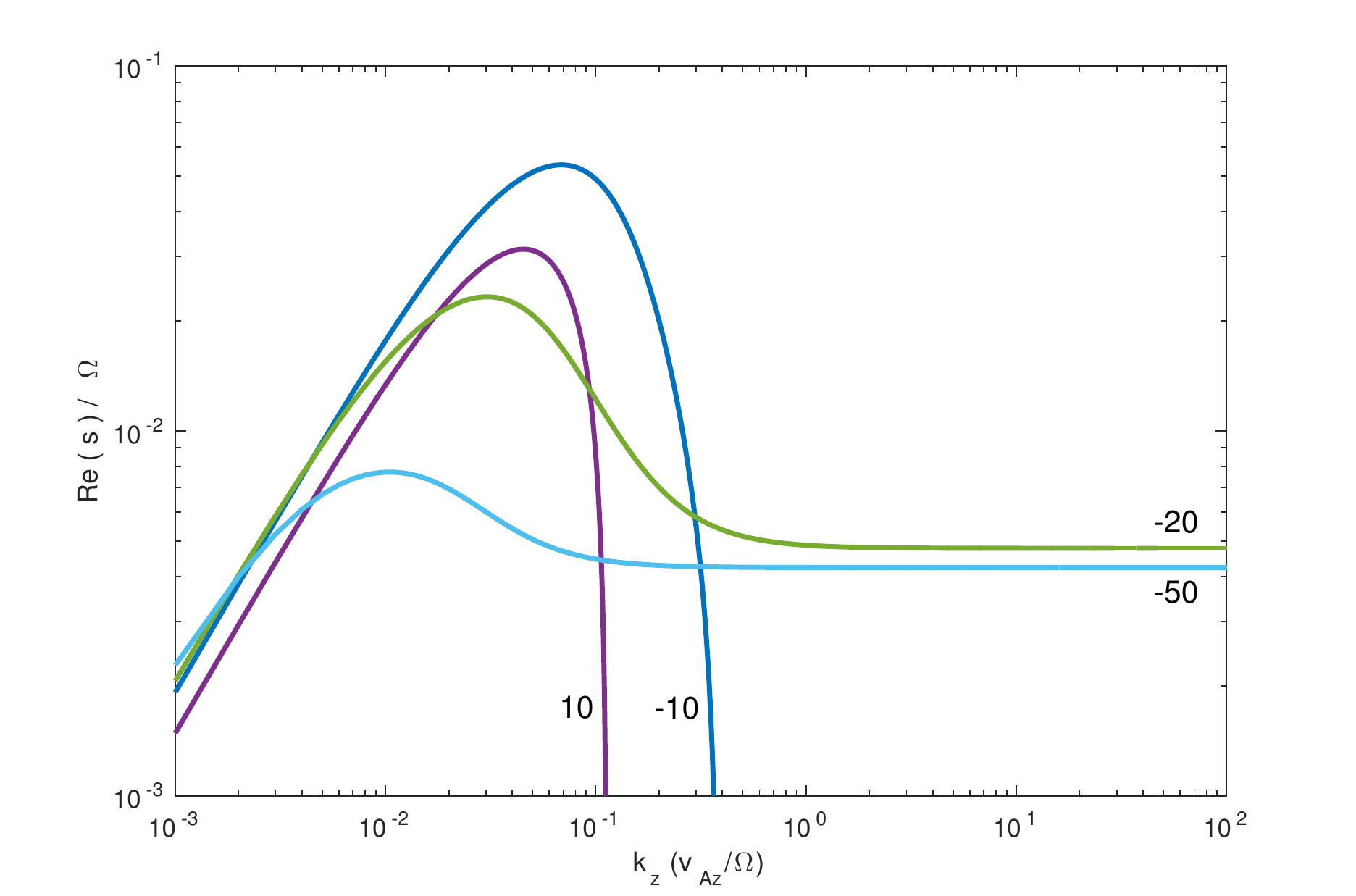}
   \caption{\label{GR2} The real part of the MRI growth rate $s$ as a
     function of vertical wavenumber $k_z$. Here $B_y/B_z=5$, $\nu_i=100$ and
     radial modes are plotted of different $k_x$. The four curves
     correspond to $k_x/k_z=-50,\,-20,\,-10,$ and 10. }
\end{figure}

Modes exhibiting selected radial wavenumbers
are presented in Figure \ref{GR2} in the case of
$B_y/B_z =5$. We show growth rates when $k_x/k_z= -50,\,-20,\,-10,$
and 10. As pointed out in KB04, growth can persist on extremely small
scales if the mode's wavevector is oriented favourably, as in the cases
of $k_x/k_z=-20$ and $-50$. Such modes differ significantly from
channel flows, as their wavevectors are almost perpendicular.
Note, however, that their growth rates are over two orders of
magnitude smaller.

The main question, when applying these results to $\beta$ Pic, is whether
unstable modes (exhibiting respectable growth rates) can actually fit
into the vertical extent of the disc. Let us first consider the channel
modes of Fig.~\ref{GR1}. They require that $v_{Az} \lesssim c_s$, 
a criterion that can be reworked into a condition on the latent
magnetic field
\begin{equation}
B_z \lesssim 4.1\times 10^{-6} 
\left(\frac{T}{100 \,\text{K}}\right)^{1/2}\left(\frac{n}{100 \,\text{cm}^{-3}}\right)^{1/2}\text{G}.
\end{equation} 
Using the estimates of Section 2, we find that
at 10 AU, the vertical field must be
less than about $10^{-4}$ G, while at 100 AU it must be less than
$\sim 2\times 10^{-6}$ G. It is clear that relatively weak fields stabilise the
MRI, owing to the very low densities of the debris-disc gas. 
But unfortunately the magnetic environment of
$\beta$ Pic, or analogous debris discs, is difficult to constrain. As a
guide, the
Solar System's
interplanetary medium varies between $10^{-6}$ G and $10^{-5}$
\citep[e.g.][]{2005JGRA..11010212J,2012ApJ...744...51B}. Of course, $\beta$ Pic's magnetic
environment may differ markedly, it being forced by a young A-type
star. Yet it is likely that at 100 AU channel modes are at best
marginally stable to MRI channel modes, and quite possibly stable.

On the other hand, radial modes are far more difficult to stabilise,
as illustrated in Figure \ref{GR2}. Their growth
rates, however are $\lesssim 0.01\Omega$, yielding a linear 
growth time of $\gtrsim 10^5$ years. This is not only rather slow, but
the ensuing turbulence may be insufficiently vigorous to supply the $\alpha$
values necessary to match the observations. 

Finally note that this issue is complicated by the fact that once the MRI
starts in one part of the disc, it can reconfigure the
magnetic field at its current radial 
location and possibly at other locations as well, given sufficient
time. Also the presence of an embedded planet's magnetic field may
also complicate the picture. The MRI will certainly be quenched in the
vicinity of a planet with a strong internal dynamo, equivalent to
Jupiter's or Earth's.

\section{Discussion}

The previous section indicates that according to linear theory
the MRI may arise at certain radii, but it is not guaranteed
throughout the disc, at least in the form
familiar from simulations of ideal MHD. The gas is too dilute, 
in particular at larger radii $\gtrsim 100$ AU. At these radii the
more vigorous channel modes are possibly suppressed and activity
proceeds via the slower growing radial modes. Whether we obtain the
classical MRI or not comes down to the strength of the ambient
magnetic field, which is poorly constrained.  

However, let us suppose the latent field is sufficiently weak for the
linear MRI to begin. What are
the properties of its nonlinear saturated state? We are especially
interested in how efficiently the ensuing turbulence can transport
angular momentum. A number of numerical studies have been conducted
probing MRI-induced turbulence with and without ambipolar diffusion.
We now summarise these.

Simulations in purely local boxes have been performed by \citet{1998ApJ...501..758H} 
and \citet{2011ApJ...736..144B}. These show that when $\nu_i\sim
100$ the turbulence differs little from ideal MHD. Of note is that
$\alpha$ increases as the plasma beta decreases; as we expect $\beta$
to be low in the disc (because $n$ is small), this hints that the
accretion rate could be as high as 0.1, as required \citep{kral16}.
Indeed, recent \emph{ideal} MHD simulations using vertically stratified boxes
show $\alpha$ as high as 1 when the plasma beta of the imposed
vertical field dips below 100 \citep{2016MNRAS.457..857S}. 
The disc, in these cases, becomes
magnetically dominated, with its vertical structure controlled
by magnetic pressure. It may be that debris disc gas, being so dilute,
falls into this turbulent and magnetically dominated regime and thus
exhibits the large alphas required by modelling and observations. 

On the other hand, work incorporating both non-ideal MHD and vertically
stratified boxes has focussed on
the specific conditions of protoplanetary
discs \citep{2013ApJ...767...30B,2013ApJ...764...66S,2013ApJ...775...73S}. These
simulations are strongly influenced by the vertical profiles of the
ionisation fraction, which vary significantly from the surface to the
midplane, where $\nu_i$ can dip to very low levels indeed. It is
difficult to directly apply these results to the gas swathing debris
discs, but they do sketch out possible behaviours that debris discs
may exhibit. Of greatest interest is the suppression of the MRI and
the launching of a magnetocentrifugal wind once the imposed vertical 
magnetic field grows too strong. 
Though mass is lost via the wind it also leads to significant angular
momentum transport and hence radial accretion
\citep[e.g.,][]{2012MNRAS.423.1318O,2013A&A...550A..61L}.
 The effective $\alpha$
associated with this process may provide another route by which debris
disc gas diffuses to smaller radii, as suggested by the modelling of
observations (Section 2). In fact, given the expected lower betas at
outer radii it is tempting to posit that the outer disc suffers a wind, while the
inner radii undergo MRI turbulent accretion.  

Obviously, many of these issues could be settled by dedicated
vertically stratified simulations using vertical ionisation profiles
appropriate for debris discs (as opposed to protoplanetary discs). 
Such computations could establish what kind of MHD behaviour develops at
different radii in a model of $\beta$ Pic, especially when the plasma
beta is near (or less than) one. 
In this case, we may be able to determine whether radial MRI modes or disc winds
induce sufficient transport to match the observations.

\section{Conclusion}\label{conclusion}
 In this paper, we tested the linear stability of the gaseous envelope
 of a debris disc to the MRI.
 After giving a general theoretical framework to be applied to any debris disc, we
 focused on $\beta$ Pic, which is the best characterised system.
 We showed that amongst the non ideal terms, the Ohmic dissipation and
 Hall effect can be neglected, while
 ambipolar diffusion is only moderately important in most parts of the disc. We
 show that given an ambient field less than a $\mu$G
  the MRI will grow in most parts of the gas disc ($\lesssim$ 200AU).
 It is possible that even if stabilised by magnetic tension a
 magnetocentrifugal wind could be launched that would transport
 angular momentum in place of the MRI.
 The presence of a strong embedded magnetic field, such as around
 the planet $\beta$ Pic b, would
 complicate the picture and may quench the MRI at the location of
 the planet.
 The model used in this paper could be used to assess the efficiency of
 MRI in other debris discs
 where the ionisation fraction, gas and electron densities might
 differ.

 We propose some future work
 to be able to better quantify the MRI or MHD processes (winds)
 at different radii through the disc. This could
 be done undertaking vertically stratified simulations of debris disc
 gas.

\section*{Acknowledgments}
We thank the referee for his/her valuable comments concerning the dust component. QK acknowledge support from the European Union through ERC grant
number 279973.
HNL is partially supported by STFC grant ST/L000636/1.

\label{lastpage}


\begin{thebibliography}{}
\bibitem[Bai(2011)]{2011ApJ...739...50B} Bai, X.-N.\ 2011, \apj, 739, 50 
\bibitem[Bai \& Stone(2011)]{2011ApJ...736..144B} Bai, X.-N., \& Stone, J.~M.\ 2011, \apj, 736, 144 
\bibitem[Bai \& Stone(2013a)]{2013ApJ...767...30B} Bai, X.-N., \& Stone, J.~M.\ 2013, \apj, 767, 30 
\bibitem[Bai \& Stone(2013b)]{2013ApJ...769...76B} Bai, X.-N., \& Stone, J.~M.\ 2013, \apj, 769, 76
\bibitem[Balbus \& Hawley(1998)]{1998RvMP...70....1B} Balbus, S.~A., \& Hawley, J.~F.\ 1998, Reviews of Modern Physics, 70, 1 
\bibitem[Bell et al.(1997)]{1997ApJ...486..372B} Bell, K.~R., Cassen, P.~M., Klahr, H.~H., \& Henning, T.\ 1997, \apj, 486, 372
\bibitem[Blaes \& Balbus(1994)]{1994ApJ...421..163B} Blaes, O.~M., \& Balbus, S.~A.\ 1994, \apj, 421, 163 
\bibitem[Burlaga \& Ness(2012)]{2012ApJ...744...51B} Burlaga, L.~F., \& Ness, N.~F.\ 2012, \apj, 744, 51  
\bibitem[Carciofi et al.(2012)]{2012ApJ...744L..15C} Carciofi, A.~C., Bjorkman, J.~E., Otero, S.~A., et al.\ 2012, \apjl, 744, L15
\bibitem[Cataldi et al.(2014)]{2014A&A...563A..66C} Cataldi, G., Brandeker, A., Olofsson, G., et al.\ 2014, \aap, 563, A66 
\bibitem[Czechowski \& Mann(2007)]{2007ApJ...660.1541C} Czechowski, A., \& Mann, I.\ 2007, \apj, 660, 1541 
\bibitem[Dent et al.(2014)]{2014Sci...343.1490D} Dent, W.~R.~F., Wyatt, M.~C., Roberge, A., et al.\ 2014, Science, 343, 1490 
\bibitem[Draine \& Sutin(1987)]{1987ApJ...320..803D} Draine, B.~T., \& Sutin, B.\ 1987, \apj, 320, 803
\bibitem[Draine(2011)]{2011piim.book.....D} Draine, B.~T.\ 2011, Physics of the Interstellar and Intergalactic Medium by Bruce T.~Draine.~Princeton University Press, 2011.~ISBN: 978-0-691-12214-4
\bibitem[Ferland et al.(2013)]{2013RMxAA..49..137F} Ferland, G.~J., Porter, R.~L., van Hoof, P.~A.~M., et al.\ 2013, \rmxaa, 49, 137 
\bibitem[Fern{\'a}ndez et al.(2006)]{2006ApJ...643..509F} Fern{\'a}ndez, R., Brandeker, A., \& Wu, Y.\ 2006, \apj, 643, 509 
\bibitem[Freudling et al.(1995)]{1995A&A...301..231F} Freudling, W., Lagrange, A.-M., Vidal-Madjar, A., Ferlet, R., \& Forveille, T.\ 1995, \aap, 301, 231 
\bibitem[Goldreich \& Lynden-Bell(1965)]{1965MNRAS.130...97G} Goldreich, P., \& Lynden-Bell, D.\ 1965, \mnras, 130, 97 
\bibitem[Grigorieva et al.(2007)]{2007A&A...475..755G} Grigorieva, A.,
  Th{\'e}bault, P., Artymowicz, P., \& Brandeker, A.\ 2007, \aap, 475,
  755 
\bibitem[Guilet et al.(2015)]{2015MNRAS.447.3992G} Guilet, J.,
  M{\"u}ller, E., Janka, H.-T., \mnras, 447, 3992
\bibitem[Hawley \& Stone(1998)]{1998ApJ...501..758H} Hawley, J.~F., \& Stone, J.~M.\ 1998, \apj, 501, 758 
\bibitem[Hayashi(1981)]{1981PThPS..70...35H} Hayashi, C.\ 1981, Progress of Theoretical Physics Supplement, 70, 35 
\bibitem[Heap et al.(2000)]{2000ApJ...539..435H} Heap, S.~R., Lindler, D.~J., Lanz, T.~M., et al.\ 2000, \apj, 539, 435 
\bibitem[Jackman et al.(2005)]{2005JGRA..11010212J} Jackman, C.~M., Achilleos, N., Bunce, E.~J., et al.\ 2005, Journal of Geophysical Research (Space Physics), 110, A10212 
\bibitem[King et al.(2007)]{2007MNRAS.376.1740K} King, A.~R., Pringle, J.~E., \& Livio, M.\ 2007, \mnras, 376, 1740
\bibitem[Kral et al.(2013)]{2013A&A...558A.121K} Kral, Q., Th{\'e}bault, P., \& Charnoz, S.\ 2013, \aap, 558, A121  
\bibitem[Kral et al.(2016)]{kral16} Kral, Q., Wyatt, M., Carswell, R., Pringle, J., Matr\`a, L., \& Juh\'asz, A. \ 2016, \MNRAS, subm.
\bibitem[Kunz \& Balbus(2004)]{2004MNRAS.348..355K} Kunz, M.~W., \& Balbus, S.~A.\ 2004, \mnras, 348, 355
\bibitem[Latter et al.(2015)]{2015MNRAS.453.3257L} Latter, H.~N., Fromang, S., \& Faure, J.\ 2015, \mnras, 453, 3257
\bibitem[Lecavelier des Etangs et al.(2001)]{2001Natur.412..706L}
  Lecavelier des Etangs, A., Vidal-Madjar, A., Roberge, A., et al.\
  2001, \nat, 412, 706
\bibitem[Lesur et al.(2013)]{2013A&A...550A..61L} Lesur, G.,
  Ferreira, J., Ogilvie, G.~I., 2013, \aap, 550, 61
\bibitem[Lesur et al.(2014)]{2014A&A...566A..56L} Lesur, G., Kunz,
  M.~W., \& Fromang, S.\ 2014, \aap, 566, A56 
\bibitem[Ogilvie (2012)]{2012MNRAS.423.1318O} Ogilvie, G.~I., 2012,
  \mnras, 423, 13180
\bibitem[Olofsson et al.(2001)]{2001ApJ...563L..77O} Olofsson, G., Liseau, R., \& Brandeker, A.\ 2001, \apjl, 563, L77 
\bibitem[Roberge et al.(2006)]{2006Natur.441..724R} Roberge, A., Feldman, P.~D., Weinberger, A.~J., Deleuil, M., \& Bouret, J.-C.\ 2006, \nat, 441, 724 
\bibitem[Roberge et al.(2014)]{2014ApJ...796L..11R} Roberge, A.,
  Welsh, B.~Y., Kamp, I., Weinberger, A.~J., \& Grady, C.~A.\ 2014,
  \apjl, 796, L11 
\bibitem[Salvesen et al.(2016)]{2016MNRAS.457..857S} Salvesen, G., Simon, J.~B., Armitage, P.~J., Begelman, M.~C.\ 2016, \mnras, 457, 857
\bibitem[Sano et al.(2004)]{2004ApJ...605..321S} Sano, T., Inutsuka, S.-i., Turner, N.~J., \& Stone, J.~M.\ 2004, \apj, 605, 321
\bibitem[Simon et al.(2009)]{2009ApJ...690..974S} Simon, J.~B., Hawley, J.~F., \& Beckwith, K.\ 2009, \apj, 690, 974 
\bibitem[Simon et al.(2013a)]{2013ApJ...764...66S} Simon, J.~B., Bai, X.-N., Stone, J.~M., Armitage, P.~J., \& Beckwith, K.\ 2013, \apj, 764, 66 
\bibitem[Simon et al.(2013b)]{2013ApJ...775...73S} Simon, J.~B., Bai, X.-N., Armitage, P.~J., Stone, J.~M., \& Beckwith, K.\ 2013, \apj, 775, 73 
\bibitem[Simon et al.(2015)]{2015MNRAS.454.1117S} Simon, J.~B., Lesur, G., Kunz, M.~W., \& Armitage, P.~J.\ 2015, \mnras, 454, 1117 
\bibitem[Shakura \& Sunyaev(1973)]{1973A&A....24..337S} Shakura, N.~I., \& Sunyaev, R.~A.\ 1973, \aap, 24, 337
\bibitem[Zagorovsky et al.(2010)]{2010ApJ...720..923Z} Zagorovsky, K., Brandeker, A., \& Wu, Y.\ 2010, \apj, 720, 923 

\end{thebibliography}
\end{document}